\documentclass[12pt]{article}
\usepackage {amsmath}
\usepackage[dvips]{graphicx}
\usepackage{amssymb}
\usepackage{cite}
\newcommand{\be}{\begin{equation}}
\newcommand{\ee}{\end{equation}}
\newcommand{\bear}{\begin{eqnarray}}
\newcommand{\eear}{\end{eqnarray}}

\newcommand{\lapproxeq}{\lower .7ex\hbox{$\;\stackrel{\textstyle  
<}{\sim}\;$}} 
\newcommand{\gapproxeq}{\lower .7ex\hbox{$\;\stackrel{\textstyle  
>}{\sim}\;$}} 
\newcommand{\stackdown}[2]{\lower 1.4ex\hbox{$\;\stackrel{\textstyle{#1}}  
{\scriptstyle{#2}}\;$}}
\newcommand{\beq}{\begin{equation}} 
\newcommand{\eeq}{\end{equation}} 

\newcommand{\bea}{\begin{eqnarray}}
\newcommand{\eea}{\end{eqnarray}}

\makeatletter 
\def\slash{\@ifnextchar[{\fmsl@sh}{\fmsl@sh[0mu]}} 
\def\fmsl@sh[#1]#2{%
  \mathchoice 
    {\@fmsl@sh\displaystyle{#1}{#2}}%
    {\@fmsl@sh\textstyle{#1}{#2}}%
    {\@fmsl@sh\scriptstyle{#1}{#2}}%
    {\@fmsl@sh\scriptscriptstyle{#1}{#2}}} 
\def\@fmsl@sh#1#2#3{\m@th\ooalign{$\hfil#1\mkern#2/\hfil$\crcr$#1#3$}} 
\makeatother 

\begin{document} 
\baselineskip=18pt 

\begin{titlepage}
\rightline{ACT-04-06, MIFP-06-12}
\rightline{UOA-NPPS/BSM-06-02}
\vspace{1cm}
\begin{center}
{\bf {\large Dissipative Liouville Cosmology: A Case Study }}\\
\vspace{0.3in}
\end{center}
\begin{center}
\vskip 0.05in
{\bf G.A.~Diamandis }$^{1}$, {\bf B.~C.~Georgalas}$^{1}$, 
{\bf A.~B.~Lahanas}$^{1}$,~ {\bf N.E.~Mavromatos}$^{2}$,
and {\bf D.V.~Nanopoulos}$^{3}$

\vskip 0.5in
{\it
$^1${University of Athens, Physics Department, 
Nuclear and Particle Physics Section,  GR157 71, 
Athens, Greece}\\
$^2${King's College London, Physics Department, 
Theoretical Physics, 
Strand WC2R 2LS, UK}\\
$^3${George P.\ and Cynthia W.\ Mitchell Institute for
Fundamental
Physics, \\ Texas A\&M
University, College Station, TX 77843, USA; \\
Astroparticle Physics Group, Houston
Advanced Research Center (HARC),
Mitchell Campus,
Woodlands, TX~77381, USA; \\
Academy of Athens,
Division of Natural Sciences, 28~Panepistimiou Avenue,
Athens 10679,
Greece}}\\

\vspace{1cm}
{\bf Abstract}
\end{center}

We consider solutions of the cosmological equations
pertaining to a dissipative, dilaton-driven 
off-equilibrium Liouville Cosmological model,
which may describe the effective field theoretic limit of a non-critical
string model of the Universe. The non-criticality may be the result
of an early-era catastrophic cosmic event, such as a big-bang, 
brane-world collision {\it etc}.
The evolution of the various cosmological parameters of the model 
are obtained, and the effects of the dilaton and off-shell Liouville terms,
including briefly those on relic densities, 
which distinguish the model from conventional cosmologies,  
are emphasised.

\vspace{0.5cm}
\leftline{May 2006}
\end{titlepage}

The current astrophysical data~\cite{snIa,wmap,Upadhye:2004hh,deceldata} 
are capable of placing stringent 
constraints on the nature of the dark energy, whose equation of state
may be determined by means of an appropriate global fit.
Most of the analyses so far are based
on effective four-dimensional Robertson-Walker Universes, which
satisfy on-shell dynamical equations of motion of the
Einstein-Friedmann form. Even in modern approaches to brane cosmology,
which are described by equations that deviate during early eras of the
Universe from the standard Friedmann equation (which is linear in the
energy density), the underlying dynamics is assumed to be of classical
equilibrium (on-shell) nature, in the sense that it satisfies a set of
equations of motion derived from the appropriate minimisation of an
effective space-time Lagrangian.

However, cosmology may not be an entirely classical equilibrium
situation \cite{aben,emnw}.  The initial Big Bang or other catastrophic
cosmic event, such as the collision of two brane worlds 
in the modern context of brane theories~\cite{polchinski}, 
which led to the initial rapid expansion of the
Universe, may have caused a significant departure from classical
equilibrium dynamics in the early Universe, whose signatures may still
be present at later epochs including the present era. 
In \cite{emnw,diamandis}
there have been proposed specific models for the cosmological
dark energy which are of this type, being associated with a rolling
dilaton field that is a remnant of this non-equilibrium phase,
described by a generic non-critical string
theory~\cite{aben,ddk,emn}. The basic ingredient of this approach is the identification 
of target time with 
a local, dynamical (irreversible) renormalization group
scale on the world sheet of the string~\cite{gsw,polchinski}, 
being representing  
by the so-called Liouville mode~\cite{aben,emn}.
The consistency of the approach is guaranteed by the existence
of solutions to the pertinent equations (``Liouville conditions'') 
for the various 
background target-space fields over which the non-critical 
Liouville-dressed~\cite{ddk} $\sigma$-model propagates. The latter
express the restoration of conformal invariance conditions
after Liouville dressing. We call this scenario `Q-cosmology' \cite{emnw}.

It must be stressed that this 
Q-cosmology is physically very different 
from standard dilaton cosmologies in critical strings~\cite{gasperini},
where on-shell equations of motion for the background fields
are satisfied. Our 
non-equilibrium, non-classical theory {\it is not described by
the equations of motion derived by extremising an effective space-time
Lagrangian}. One must use a more general formalism
to make predictions that can be confronted with the current data. The
approach we favour is formulated in the context of string/brane
theory~\cite{polchinski,gsw}, the best candidate theory of quantum
gravity to date. Our approach is based on non-critical (Liouville)
strings~\cite{aben,ddk,emn}, which offer a mathematically consistent
way of incorporating time-dependent backgrounds in string theory.

The basic idea behind such non-critical Liouville strings is the
following. Usually, in string perturbation theory, the target space
dynamics is obtained from a stringy $\sigma$-model~\cite{gsw} that
describes the propagation of strings in classical target-space background
fields, including the space-time metric itself.  Consistency of the
theory requires conformal invariance on the world sheet, in which case
the target-space physics is independent of the scale characterising
the underlying two-dimensional dynamics. These conformal invariance
conditions lead to a set of target-space equations for the various
background fields, which correspond to the Einstein/matter equations
derived from an appropriate low-energy effective action that is
invariant under general coordinate transformations. Unfortunately, one
cannot incorporate in this way time-dependent cosmological backgrounds
in string theory, since, to low orders in a perturbative expansion in
the Regge slope $\alpha '$, the conformal invariance condition for the
metric field would require a Ricci-flat target-space manifold, whereas
a cosmological background necessarily has a non-vanishing Ricci
tensor.

To be able to describe a time-dependent
cosmological background in string theory, the authors of~\cite{aben}
suggested that a non-trivial r\^ole should be played by a
time-dependent dilaton background.  This approach leads to strings
living in numbers of dimensions different from the customary critical
number, and was in fact the first physical application of
non-critical strings~\cite{ddk}. The approach of~\cite{aben}
was subsequently extended~\cite{emn}, \cite{emnw},\cite{diamandis} to incorporate off-shell
quantum effects and non-conformal string backgrounds describing other
non-equilibrium cosmological situations, including catastrophic cosmic
events such as the collision of two brane worlds, \emph{etc.}.

In a recent work~\cite{mitsou} 
we have discussed fits of such non-critical Liouville 
cosmological dark energy models 
to the available 
data on high-redshift supernovae~\cite{snIa} 
and baryon oscillations~\cite{baryon}. 
It was found in that analysis that 
a simple
parametrisation of the full version of the Q-cosmology 
model that includes
off-shell effects fits the data very well. 
Specifically, we have shown that, 
under certain approximations, which allowed 
for an analytic solution of the pertinent
Liouville conditions, the Hubble parameter of this model, 
$H(z)$, where $z$ is the redshift,  
can be expressed in the form:
\begin{eqnarray}
H(z) &=& H_0 \left( {\Omega }_3 (1 + z)^3 + {\Omega }_\delta (1 +
z)^\delta + {\Omega}_2 (1 + z)^2 \right)^{1/2}~,\nonumber \\
\label{formulaforfit_txt}
\end{eqnarray}
with the subscript $0$ denoting present-day values ($z = 0$) and
\begin{equation}
\label{sumomegas_txt}
{\Omega}_3
+ {\Omega}_{\delta} + {\Omega}_2 = 1~,
\end{equation}
The exponent $\delta$  was treated in \cite{mitsou} as a fitting parameter. 

The basic parameter used to fit the supernova data
is the luminosity distance $d_L = c(1 + z) \int_0^z \frac{dz'}{H(z')}$,
which is connected to the apparent magnitude of the supernovae. 
The data seem to favour the value $\delta \sim 4$, which can also
be explained theoretically.
As discussed in \cite{mitsou}, the ``dust''-like contributions,  
$\Omega_3$, do not merely represent
ordinary matter effects, but also receive contributions from the
dilaton dark energy. In fact, the sign of $\Omega_3$ depends on
details of the underlying theory, and it could even be negative. For
instance, as argued in \cite{mitsou}, 
string loop corrections could lead to a negative $\Omega_3$.
In addition, 
Kaluza-Klein graviton modes in certain brane-inspired
models~\cite{kaluza} also yield negative dust contributions. In a similar
vein, the exotic contributions scaling as $(1+z)^\delta$ are affected
by the off-shell Liouville terms of Q-cosmology. It is because of the
similar scaling behaviours of dark matter and dilaton dark energy that
we reverted to the notation $\Omega_i$, $i=2,3,\delta$
in~(\ref{formulaforfit_txt}). More generally, one could have included
a cosmological constant $\Omega_\Lambda$ contribution
in~(\ref{formulaforfit_txt}), which may be induced in certain
brane-world inspired models. We did not do so in \cite{mitsou}, 
as our
primary interest is to fit Q-cosmology
models~\cite{emnw,diamandis}, which are characterised by
dark energy densities that relax to zero asymptotically 
in cosmic time.

Some remarks are now in order. First, we stress that 
the above formulae are valid for late eras, such as the ones
pertinent to the supernova and other data ($0 \le z \le 2$) 
that we used
in \cite{mitsou}. Moreover, 
in the context of Q-cosmology,  
the form (\ref{formulaforfit_txt}), 
can only be obtained  
after making a series of approximations, which may not always 
be valid, as already stressed in \cite{mitsou}.
It is the purpose of this paper to present a more complete, 
numerical 
treatment of the Liouville conditions of a case study in Q-cosmology, 
and derive the behaviour of the various cosmological 
parameters, including $H(z)$, with the cosmic time (or, equivalently, $z$).
For the interested reader, the details and terminology of Q-cosmology 
can be found in the relevant literature~\cite{aben,emnw,emn}. 

We commence our analysis 
by considering a Liouville-dressed 
non-critical $\sigma$-model propagating in cosmological 
dilaton and graviton backgrounds. After the identification
of the target time with the Liouville mode~\cite{emn},\cite{emnw},\cite{diamandis}
the relevant Liouville equations 
in the Einstein frame \cite{aben}, i.e. 
in the frame where the scalar curvature in the (off-shell) target space
effective action assumes the canonical Einstein form to leading order
in the Regge-slope $\alpha '$ expansion,  
are given by 
\bear
&&3 \; H^2 - {\tilde{\varrho}}_m - \varrho_{\phi}\;=\; \frac{e^{2 \phi}}{2} \; \tilde{\cal{G}}_{\phi} \nonumber  \\
&&2\;\dot{H}+{\tilde{\varrho}}_m + \varrho_{\phi}+
{\tilde{p}}_m +p_{\phi}\;=\; \frac{\tilde{\cal{G}}_{ii}}{a^2} \nonumber  \\
&& \ddot{\phi}+3 H \dot{\phi}+ \frac{1}{4} \; \frac{\partial { \hat{V}}_{all} }{\partial \phi}
+ \frac{1}{2} \;( {\tilde{\varrho}}_m - 3 {\tilde{p}}_m )= 
- \frac{3}{2}\; \frac{ \;\tilde{\cal{G}}_{ii}}{ \;a^2}- \,
\frac{e^{2 \phi}}{2} \; \tilde{\cal{G}}_{\phi}  \; .\label{eqall}
\eear
As usual dots denote derivatives with respect Einstein time. 
The r.h.s of these equations
constitute a manifestation of the non-critical
string, {\it off-shell},  
behaviour. 
Such  terms are absent in critical string cosmologies, such as the models 
considered in \cite{gasperini}. 
The dependence of the off-shell terms ${\cal G}$ on the
cosmic scale factor, dilaton and (square root of the) 
central charge deficit $Q$, is as follows:
\bear && \tilde{\cal{G}}_{\phi} \;=\; e^{\;-2 \phi}\;( \ddot{\phi} -
{\dot{\phi}}^2 + Q e^{\phi} \dot{\phi}) \nonumber \\ &&
\tilde{\cal{G}}_{ii} \;=\; 2 \;a^2 \;(\; \ddot{\phi} + 3 H \dot{\phi}
+ {\dot{\phi}}^2 + ( 1 - q ) H^2 + Q e^{\phi} ( \dot{\phi}+ H )\;) \;
.  \eear
Notice the {\it dissipative} terms proportional 
to $Q\dot{\phi}$, which is responsible for the terminology
``Dissipative Cosmology'' used alternatively for Q-cosmology~\cite{emnw}. 
In these equations $q$ is the deceleration $q \equiv - \ddot{a} a /
{\dot{a}}^2$ as function of the time {\footnote{ The function
${\tilde{\cal{G}}}_{00}$, which is the $00$ component of
${\tilde{\cal{G}}}_{\mu \nu}$, is zero since the corresponding
component of the metric is constant.}}.  The variation of the
central charge deficit 
$Q$ with the cosmic time is provided by the Curci - Paffuti
equation~\cite{curci}, which expresses the renormalisability 
of the world-sheet theory. To leading order in an $\alpha '$ expansion,
which we restrict ourselves in this work, 
this equation in the Einstein frame is given 
by
\bear \frac{d \tilde{\cal{G}}_{\phi} }{d t_E} \;=\; - 6\; e^{\;-2
\phi}\;( H + \dot{\phi} ) \; \frac{ \;\tilde{\cal{G}}_{ii}}{ \;a^2} \;
.  \label{CXP} \eear
We remind the reader once more that all quantities in the above
equations refer to the Einstein frame~\cite{aben}, and derivatives are
with respect the Einstein time $t_E$ which is related to the real time
$t$ by $t=\omega^{-1} \; t_E$. $\omega$ is an arbitrary dimensionful
constant with units of inverse time.

The potential appearing above is defined by 
$\; {\hat{V}}_{all}= 2 Q^{\;2} \exp{\;( 2 \phi )}+V \;$ where 
in order to cover more general cases we have allowed 
for a potential term in the string action $- \sqrt{-G}\;V$ in addition 
to that dependent on the central charge deficit term. Although we 
have assumed a (spatially) 
flat Universe the terms on the r.h.s., which manifest 
departure from the criticality, act in a sense like curvature terms 
as being non-zero at certain epochs. The dilaton energy 
density and pressure are 
\bear && \varrho_{\phi}\;=\;\frac{1}{2}\;( \; 2
{\dot{\phi}}^2+{\hat{V}}_{all} ) \nonumber \\
&&p_{\phi}\;=\;\frac{1}{2}\;( \; 2 {\dot{\phi}}^2-{\hat{V}}_{all} ) \;
\; .  \eear
The dilaton field is not canonically normalised in this convention and
its dimension has been set to zero. 

In the following we shall drop the
explicit potential term $V$ and keep only that term proportional to
the central charge deficit squared $Q^2$. This is the model considered
in \cite{emn,emnw,diamandis}. Note, however, that, as mentioned
previously, string loop
corrections, which contribute to $V$,
might be significant at present eras, as seems to be indicated
by the data~\cite{mitsou}. Such an issue can be answered
only after making detailed fits of the current numerical
results to the actual data, along the lines of \cite{mitsou}.
This is left for future work. 

The matter - energy density ${\tilde{\varrho}}_m $ is dimensionless
and is related to the actual density, occurring in the system of equations
where all quantities have their proper dimensions, through
$$
{\tilde{\varrho}}_m \;=\; \frac{ 8 \pi G_N}{\omega^2} \; {{\varrho}}_m \; .
$$ 
It is evident that a convenient choice for $\omega$
is $\omega=\sqrt{3} \; H_0$ where $H_0$ is the value of the Hubble
constant. With this choice ${\tilde{\varrho}}_m$ become exactly the
ratio of the actual density to the critical density in the
conventional Cosmology $\varrho_c$ usually denoted by $\Omega$. We
shall adhere to this choice of $\omega$ in the following.

All equations can be converted to the string frame which is
characterised by a string time $t_s$. Since the equations do not
explicitly depend on time we are allowed to perform a shift in the
variable $t_s$ and take the string time in such a way that $t_s=0$
corresponds to the present time. 
The reader should recall that the Einstein time is related to the
string time $t_s$ through~\cite{aben}
$$
t_E\;=\; \int_{t_{sing}}^{t_s} e^{ - \phi} dt_s + C \;.
$$
with $C$ an arbitrary constant having no physical significance either.
The lower limit of the integration designates the point of the initial
singularity where $Q$ or $\phi$ become infinite. The arbitrary
constant $C$ is usually taken to be zero so that in the Einstein frame
the origin of the time,$\; t_{Einstein}=0$, is when the singularity is
created.

In our numerical procedure we found it more convenient, although
physicswise less transparent, to work directly in the string frame. In
such an approach, all equations and boundary conditions should be
converted to string coordinates by converting derivatives with respect
$t_E$ to those with respect $t_s$ and using, at the same time, the
string cosmic scale factor related to the Robertson Walker Friedman
(RWF) cosmic scale factor $a$ by $a_s= e^{\phi} \; a$.  In this frame,
in order to solve the system of equations, one needs the initial
values of the quantities involved at some point which we take to be
$t_s=0$, that is the present time. Using the measured values of the
present-day 
Hubble constant $H_0$ and the deceleration $q_0$ today, the system has
a unique solution provided that the value of the dilaton field, the
matter-energy density and the central charge deficit $Q$ are also
given. 

As we shall demonstrate later on, $Q(t)$, which in our approach is 
in general cosmic time dependent~\cite{emn},\cite{diamandis}, 
is determined in terms of the
matter-energy density, the value of the dilaton field today, as well
as the values of $H_0, q_0$, through an algebraic equation which
follows from the system of differential equations. 
Moreover, when we convert the system of equations to a system of first order differential equations   
using as independent variable the string time, 
the differential equations, as well as the boundary conditions,
are manifestly invariant 
under a constant dilaton shift $\phi \rightarrow \phi +
k$, followed by a rescaling of the density~\footnote{We tacitly assume
that pressure and density for all matter species are related by an
equation of state so that a boundary condition for density yields
automatically a boundary condition for the pressure too. }, by
${\tilde{\varrho}}_m \rightarrow {\tilde{\varrho}}_m \; e^{2 k}$.

This observation is crucial,
since it actually implies that, if a solution with a given initial
condition is found, then another solution of the same problem can be
generated by merely shifting the dilaton and rescaling the density as
above.  
Due to this, the available parameter space is significantly
reduced.  In fact if one considers values for the matter density in the
range $a \leq {\tilde{\varrho}}_m \leq b$, then 
all values in the stripe $-\infty \leq \phi \leq +\infty \;,\;
a \leq {\tilde{\varrho}}_m \leq b$ 
of the dilaton-energy-density plane 
should be 
considered {\it a priori}. 
However, due to the above property one has actually to
consider only the line segment $ \phi=0, a \leq {\tilde{\varrho}}_m
\leq b$, $ \phi > 0, {\tilde{\varrho}}_m = a$ and $ \phi < 0,
{\tilde{\varrho}}_m = b$. Any other point in the allowed stripe can be
projected to these segments. This reduces considerably the numerical
effort. 
Moreover from the values of $\phi$, large positive values are
not allowed since they correspond to large values of the string
coupling outside the perturbative regime for which our equations
hold. Only $\phi \sim {\cal{O}} (1)$ can therefore be trusted. This
reduces the parameter space even more.

In order to obtain the evolution of the matter-energy density with
time it is convenient to use the continuity equation for matter, which
follows by combining the set of equations (\ref{eqall}),
\bear
\frac{d {\tilde{\varrho}}_m }{dt_E}+ 3 H ( {\tilde{\varrho}}_m +{\tilde p}_m) +
\frac{\dot Q}{2} \frac{\partial { \hat{V}}_{all} }{\partial Q} - 
\dot{\phi}\;({\tilde{\varrho}}_m - 3 {\tilde{p}}_m )\;=\;
6\;(H+\dot{\phi})\; \frac{ \;\tilde{\cal{G}}_{ii}}{a^2} \; . \label{contin}
\eear
To proceed further one needs to make some extra assumptions. 
First we assume that the matter-energy density is split as
${\tilde{\varrho}}_m = {\varrho}_b+{\varrho}_r+{\varrho}_{e}$: the
first term refers to as "dust'', $w_b=0$, and includes the baryonic matter 
and any other sort of matter, characterized by $w=0$, which does not feel the 
effect of the non-critical terms. 
The second term refers to radiation $w_r=\frac{1}{3}$ and the
third to an unknown sort of {\emph{exotic }} matter which is characterised by an
equation of state having a weight $w_e$ which in our analysis we
consider it to be an arbitrary parameter to be fitted by data
{\footnote{Recall that 
with the choice $\omega=\sqrt{3} H_0$ adopted 
${\varrho}_{b,r,e}$ are the matter-energy densities in units of the critical 
density.
}}. 
So far this does not sound
much like an assumption. However we further impose that the exotic
matter feels all the effect of the non-critical terms unlike the rest
of the species which follow continuity equations given by

\bear \frac{d {{\varrho}}_r }{dt_E}&+& 4 H {{\varrho}}_r \;=\;0
\nonumber \\
\frac{d {{\varrho}}_b }{dt_E}&+& 3 H {{\varrho}}_b - \dot{\phi}
{{\varrho}}_b \;=\;0 \, .
\label{coneq}
\eear
Notice that the first of these equations is the well-known 
continuity equation
for radiation, while the second differs from the ordinary
continuity equation only by the appearance of the dilaton dependent
terms and is the same as that given in \cite{gasperini}.  In the case
described above the exotic matter satisfies the equation
(\ref{contin}) and it is the only species assumed to be affected by the
non-criticality. If one does not like the concept of the exotic matter,
then (s)he can set $\varrho_{b}=0$ and let $\varrho_{e}$ play the r\^ole
of dust matter, choosing simultaneously $w_e=0$. Dark Matter or any sort 
of as yet undiscovered matter, such as possible supersymmetric 
particles or other matter species, can be included in either ${{\varrho}}_b $ or 
${{\varrho}}_e$, although it seems plausible to be accomodated within ${{\varrho}}_b $. 
This scheme covers the more general cases encountered in phenomenological 
applications.  

It is worth pointing out that,
due to the appearance of the dilaton
dependent term, 
the
density ${{\varrho}}_b $, 
does not scale with $a^{-3}$ as in the conventional
Cosmology~\cite{emnw,mitsou}.  
The exotic piece certainly does not scale as "dust'', not
only because of the dilaton term and the value of $w_e$
which may differ from zero, but also because it is affected by the
presence of the non-critical, off-shell terms ${\cal G}$.  
It should be stressed that no
cosmological constant is introduced by hand as this is provided, as a
relaxation effect, by the dilaton energy density.

Due to the dilaton dependence, ${{\varrho}}_b $ follows a continuity
equation which includes a $- \dot{\phi}\; {{\varrho}}_b $ term. This term
should be duly taken into account in calculating relic abundances and
depending on its sign may lead to over or under-production in addition
to the usual picture where changes in the relic abundances occur only
through interactions with the cosmic plasma. This holds independently
of the non-criticality. 

In order to be more specific, 
in the absence of dilaton couplings, 
the number
density {{$n$}} of particles of species {{$X$ }}, assumed to
be a Dark Matter candidate, changes according to the Boltzmann 
equation~\cite{kolb}
\begin{eqnarray}
\frac{dn}{dt}\;=\; - 3 \;H\; n\;-\;<\sigma v>\;( n^2-n^2_{eq}) \; .
\label{ordinarybol}
\end{eqnarray}
According to this equation, the relevant 
density is diluted because of the Universe expansion,
{{$H=\dot{a}/a\;>0$} }, but it also changes as a result 
of interactions, 
specified by the cross section term {{$\sigma$}}.
Then the pertinent energy density is given by {{$\varrho_X = n \; m_X
\quad $}}, 
from which we can obtain the relic density  
of
species $X$ as $\Omega_X \; h_0^2$, 
after solving the Boltzmann equation for n as a function of the 
cosmic time, $n(t)$, from the 
equilibrium epoch, before the freezing-point, to today's values.

However, in dilaton-driven non-critical string cosmologies, 
as becomes clear from our continuity equation
derived from the Liouville conditions above,
the energy-density of $X$ changes as
\begin{eqnarray}
\frac{d \varrho_X}{dt}\;=\; - 3 \;H\; \varrho_X\;+\;
{{\dot{\phi} \; \varrho_X +`` \cal{G}"}}
\label{changedens}
\end{eqnarray}
with ${\cal G}$ denoting the off-shell Liouville terms. 
{\footnote{Here, in order to cover more general 
situations, we adopt the point of view that Dark Matter feels the off-shell Liouville terms, 
that is it belongs to the exotic matter part discussed previously. Certainly one can ignore 
the effect of these terms and  treat in this respect Dark Matter like ordinary matter. }}

Comparing (\ref{changedens}) with (\ref{ordinarybol}), we 
observe that the conventional Boltzmann 
equation needs to be modified in Q-cosmology, in 
order to incorporate consistently 
the effects of the dilaton {\it dissipative} pressure $\sim \dot{\phi}$
and the non-critical terms, $`` \cal{G}"$ 
in the calculation of the relic density. 
A convenient and mathematically consistent way to include such effects 
is to modify the Boltzmann equation as follows: 
\begin{eqnarray}
\frac{dn}{dt}\;=\; - 3 \;H\; n\;-\;<\sigma v>\;( n^2-n^2_{eq}) 
{{\;+\;\dot{\phi} \; n + `` {\cal{G}} / {m_X} "}} \; \; .
\label{boltzman}
\end{eqnarray}
In this equation one should be cautious to avoid interpreting the last two terms as  
changing the number density of particles $X$. Actually they only change their energy as 
is apparent from (\ref{changedens}). Their inclusion in (\ref{boltzman}) is merely 
a handy way to include both effects, that of the change of the
number density and the energy due to the dilaton interaction, into a
single equation, the Boltzmann equation, and calculate, after solving it,
the relic density in the usual manner using the equation $\; \Omega_X
\; h_0^2= n\; m_X h^2_0$.  

The non-critical/dilaton modifications may have dramatic
consequences for SUSY predictions~\cite{nano} 
since the effect of the term
$\dot{\phi}$ is comparable to that of the expansion term proportional
to $H$, over large periods during the evolution of the Universe. 
This fact can already be seen in critical 
string dilaton-driven cosmologies~\cite{gasperini}
where the off-shell terms ${\cal G}$ are absent, but one still
considers the effects of the dilaton 
dissipative terms $\dot \phi$ in (\ref{boltzman}).

Following an otherwise standard analysis~\cite{kolb}, it can be 
shown~\cite{reliclmn} that in such a case the pertinent 
relic density is 
\begin{eqnarray} 
&& \Omega_X \; h_0^2 = \left({\rm no-dilaton~case}\right) \times
\left(\frac{{\tilde g_*}}{g_*}\right)^{1/2}\left(1 + \int_{\chi_0}^{\chi_f}
\frac{{\dot \phi}H^{-1}}{{\cal S}(\chi)}d\chi \right)~, \nonumber \\
&& \rho_\phi + \rho_{\rm matter} \equiv \frac{\pi^2}{30}T^4{\tilde g}~, 
\qquad {\cal S}(\chi ) \equiv \chi \; {\rm exp}\left( -\int_{\chi_0}^{\chi_f}
\frac{{\dot \phi}H^{-1}}{\chi^\prime}d\chi^\prime\right)  \; \; .
\label{relicomegadilaton}
\end{eqnarray}
In these equations $\chi \equiv T/m_X$ and $\chi_f$, $\chi_0$ refer to values 
at the freeze-out and today's temperature respectively. 
The subscript $\star$ indicates values at the freeze-out point,
and above we assumed that the presence of 
dilaton dissipative ``source terms'' proportional to ${\dot \phi}$ in the 
Boltzmann equation does not affect the thermal equilibrium, that defines
the effective number of degrees of freedom ${\tilde g}$. The untilded
quantities $g$ indicate parameters in the no-dilaton case. 

Moreover, the freeze-out point itself is affected by the presence
of the dilaton-source terms. In fact its value is shifted from 
the standard no-dilaton value $\chi_f^{\rm no~dil}$ as: 
\begin{eqnarray} 
\chi_f^{-1} =  {( \chi_f^{\rm no~dil})}^{-1} + \frac{1}{2}{\rm ln}
\left(\frac{{\tilde g_*}}{g_*}\right) + \int_{\chi_0}^{\chi_f}
\frac{{\dot \phi}H^{-1}}{\chi}d\chi
\end{eqnarray}

The presence of off-shell Liouville terms complicate 
the situation, since, unlike the dilaton dissipative terms considered above, 
they cannot be simply expressed 
as $\Gamma n$ source terms in the Boltzmann equation.
A complete treatment will be discussed in a future work~\cite{reliclmn},
where 
the relevant 
constraints imposed for the parameter space of the popular
supersymmetric schemes will also be addressed in a quantitative manner.

In the present work we
shall limit our discussion to particular background solutions which at
the present era describe a spatially flat and accelerating Universe as the
cosmological data suggest~\cite{snIa,wmap}.
One can solve numerically the set of equations
(\ref{eqall},\ref{contin} and \ref{CXP}) in order to observe the time
evolution of $\phi, a, Q$ and the densities $\varrho_b, \varrho_r$ at
various epochs, as already discussed. Certainly not all of these equations
are independent, as already remarked. For instance the continuity
equation (\ref{contin}) follows from the set of equations
(\ref{eqall}) so that one of them is actually redundant. 

In solving these 
equations we take as initial conditions the value of the Hubble
constant $H_0$, the deceleration parameter $q_0$, at the present time,
and also the values of the matter and radiation $\varrho_b, \varrho_r$
as well as the value of the exotic matter $\varrho_e$ and the value
for $w_e$ of its equation of state which we assumed
constant. The radiation at present era is small and its value is
known. For the other two it is reasonable to assume input values in
the range $0.05 \leq \varrho_b+\varrho_e \leq 0.3$. The initial
dilaton values can be non-vanishing but as already discussed the
range of its allowed values is considerably reduced since we are
forced to stay within the perturbation regime of the string
theory. Thus the calculational task is easier than anticipated.

It should be remarked that the initial value of $Q$ is not a free
parameter. In principle it is calculated within the framework 
of the underlying conformal field theory model. From an effective
theory point of view we adopt in this work, $Q(t)$ can be 
determined by combining appropriately the Liouville equations (\ref{eqall}).
The result is a quadratic algebraic equation
\bear 2 \;Q^2 - e^{- \phi} H \;Q + e^{- 2 \phi}\; ( \;{\dot{\phi}}^2 -
8 H^2 - 3 H \dot{\phi}+ \frac{5}{2} \varrho_b + \frac{8}{3} \varrho_r
\;+ \frac{5+w_e}{2} \varrho_e ) \;=\;0 \: \; , \nonumber \eear
which relates $Q(t)$ to the remaining fields. One can be convinced of that 
by taking the differential of this
equation which is proved to vanish if the set of equations (\ref{eqall},\ref{CXP}) are
satisfied. 

This equation can then be used to obtain the value 
$Q_0$ of the central charge $Q$ at zero string time, 
that is today, in terms of the remaining inputs. Using our numerical code we
have shown that one of the roots of this equation leads to
cosmologically sensible solutions, but the other fails to reproduce a
reasonable picture of the evolution of the Universe.

The initial values $H_0, q_0$ by themselves determine the derivative
of the dilaton field ${\dot \phi}_0$ today, as can be seen by
combining the differential equations, and hence the dilaton kinetic
energy. On the other hand from the previous equation it is obvious
that the combination $\hat{Q} = {Q} e^{\phi}$ is uniquely determined
for given densities and $H_0, q_0$. As an effect the potential energy
$2\;Q^2 e^{2 \phi}$ of the dilaton energy is also fixed. Therefore the
dilaton energy and pressure today do not depend on the particular
dilaton inputs and they are set once the Hubble constant, the
deceleration and the today's values of the densities as well as the
value of $w_e$ are given.

In the following we present the results of our numerical analysis
taking $H_0^{-1}=13.4 \times 10^{9} \; yrs$, corresponding to a
rescaled Hubble constant $h_0=0.73$, $q_0=-0.61$, $\varrho_r= 5.0
\times 10^{-5}$, which is today's value for the radiation density in
units of the critical density, and various initial values for the
"dust" and "exotic" densities, $\varrho_b, \varrho_e$.  As already
remarked, the value of $w_e$ is also a free parameter which in our
analysis we allow it to vary.  

\begin{figure}[H] 
\begin{center}
\includegraphics[width=7cm]{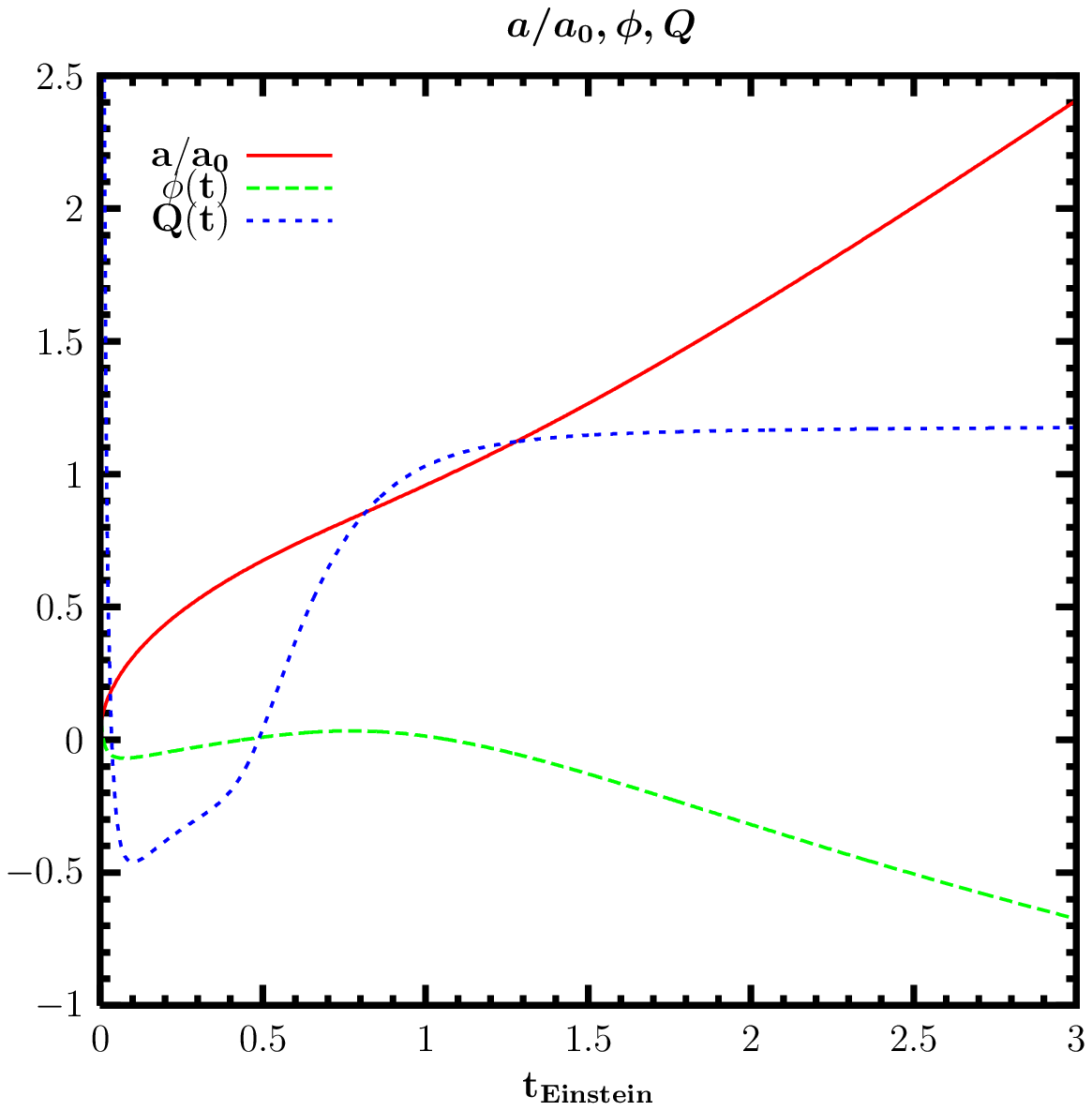}
\includegraphics[width=7cm]{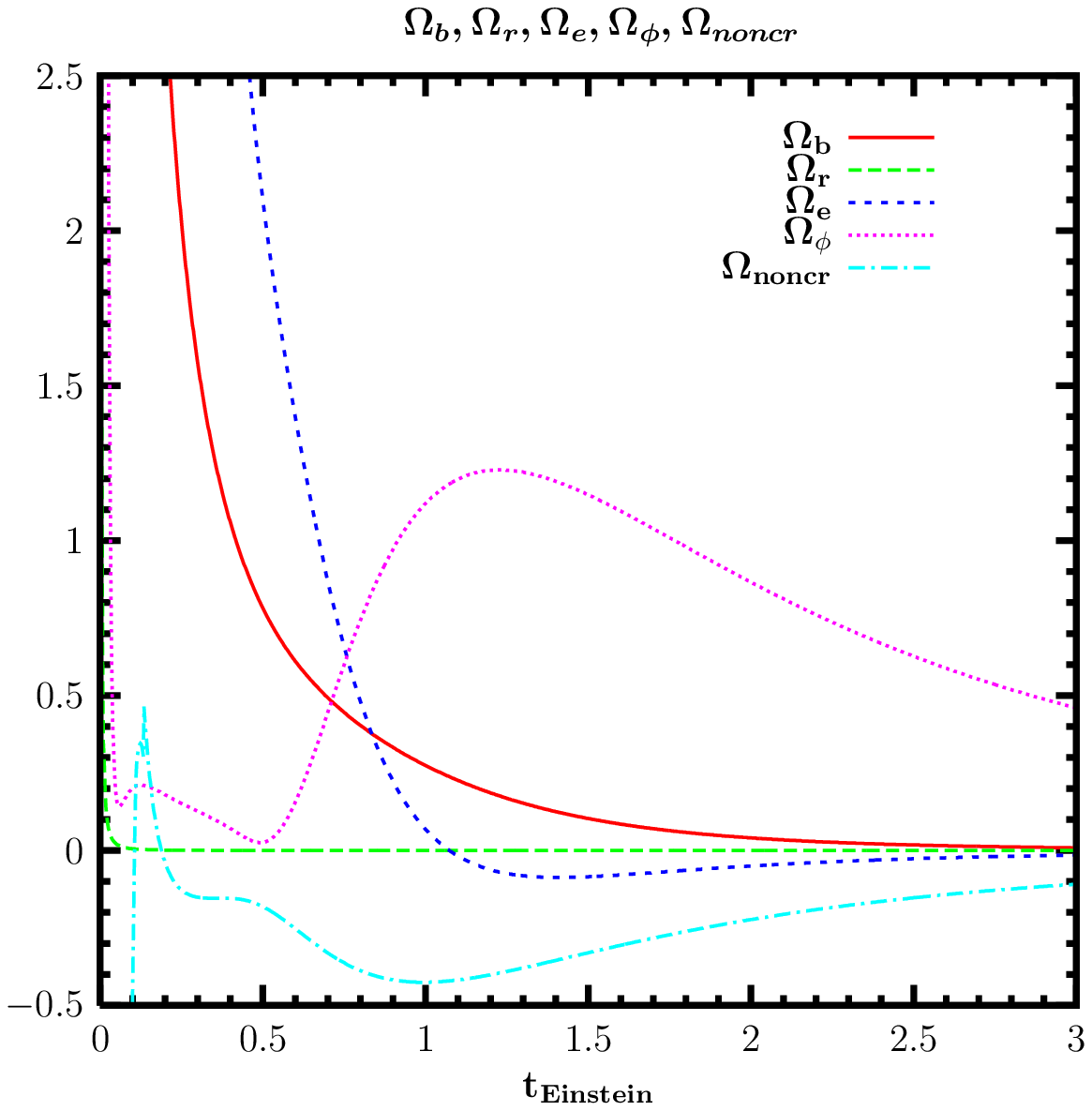}
\end{center}
\caption[]{ Left panel: The dilaton $\phi$, the (square root of the) 
central charge deficit
$Q$ and the ratio $a/a_0$ of the cosmic scale factor as functions of
the Einstein time $t_{Einstein}$. The present time is located where
$a/a_0=1$ and in the figure shown corresponds to $t_{today} \simeq
1.07$. The input values for the densities are $\rho_b=0.238,
\rho_{e}=0.0$ and $w_e$ is 0.5. The dilaton value today is taken
$\phi=0.0$ .  Right panel: The values of $\Omega_i \equiv \rho_i/\rho_c$
for the various species as functions of $t_{Einstein}$.  }
\label{fig12}  
\end{figure}

\begin{figure}[H] 
\begin{center}
\includegraphics[width=7cm]{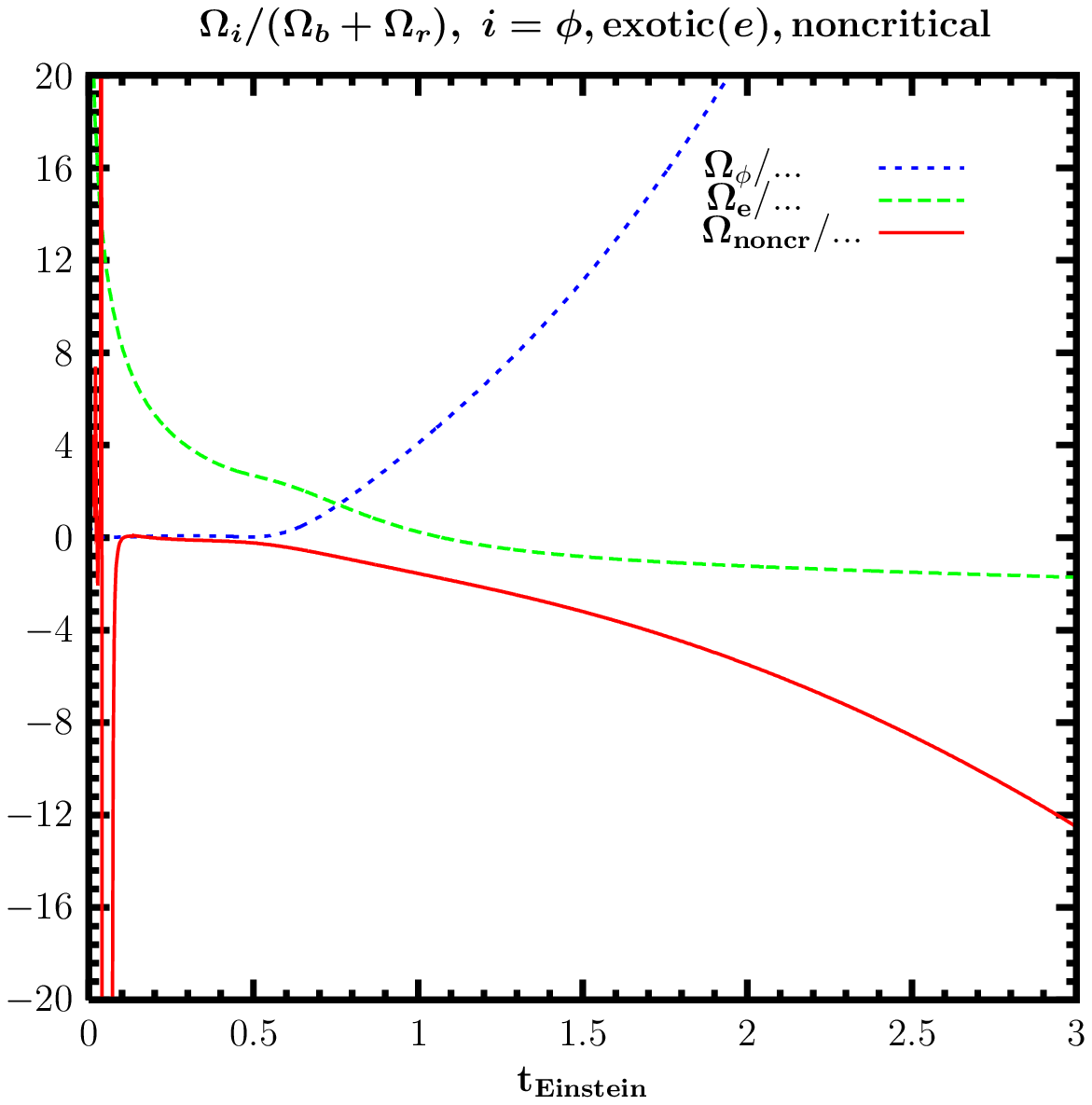}
\includegraphics[width=7cm]{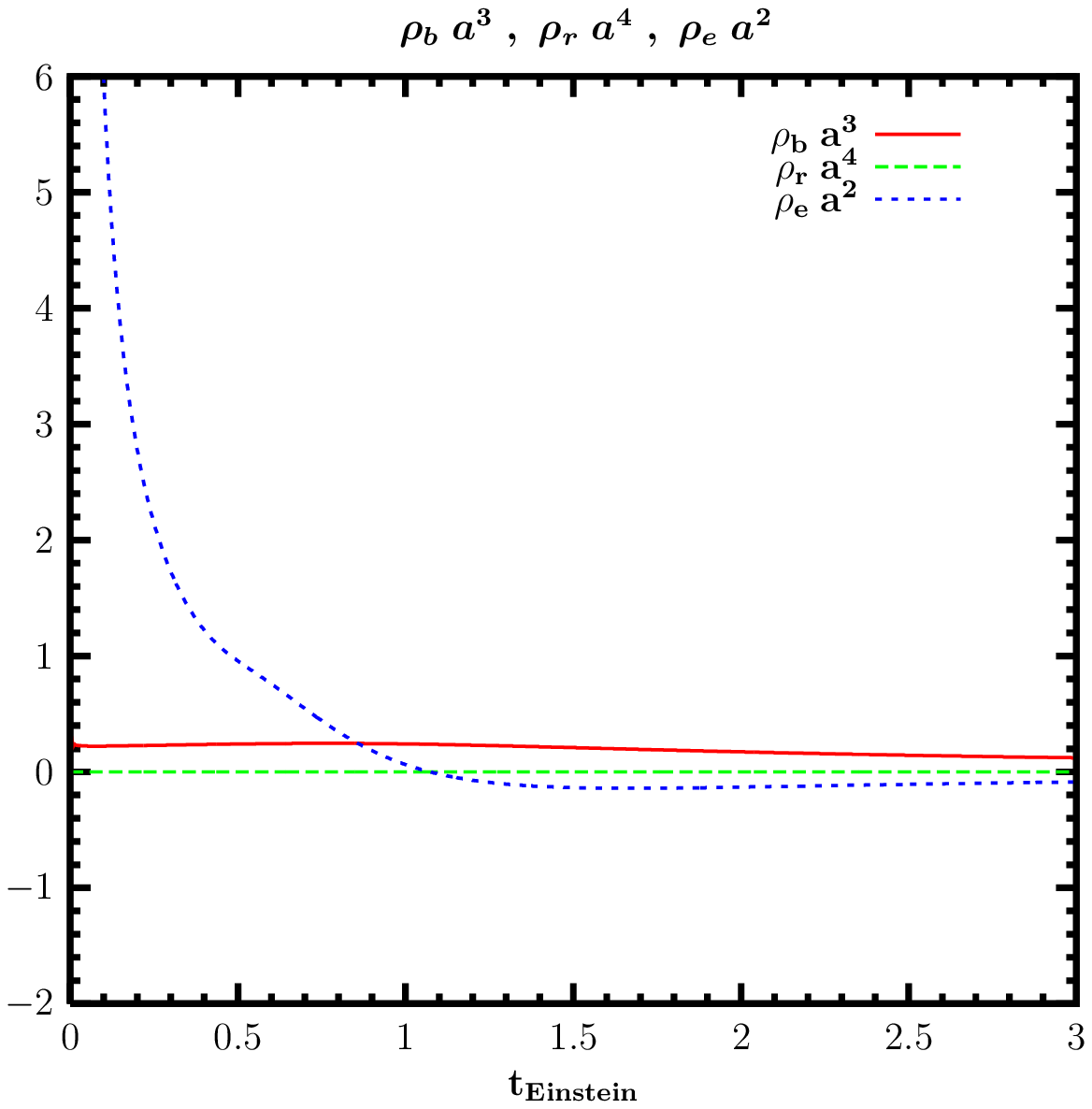}
\end{center}
\caption[]{ Left panel: Ratios of $\Omega$'s for the dilaton ($\phi$), exotic
matter ($e$) and the non-critical terms ("noncrit'') to the sum of "dust'' and radiation 
$\Omega_b + \Omega_r$ densities. \\
Right panel: The quantities $\rho_b \;a^3$, for "dust", $ \;\rho_r
\;a^4$ and $\rho_{e} \;a^2$ as functions of $t_{Einstein}$.  }
\label{fig34}  
\end{figure}

\begin{figure}[H] 
\begin{center}
\includegraphics[width=7cm]{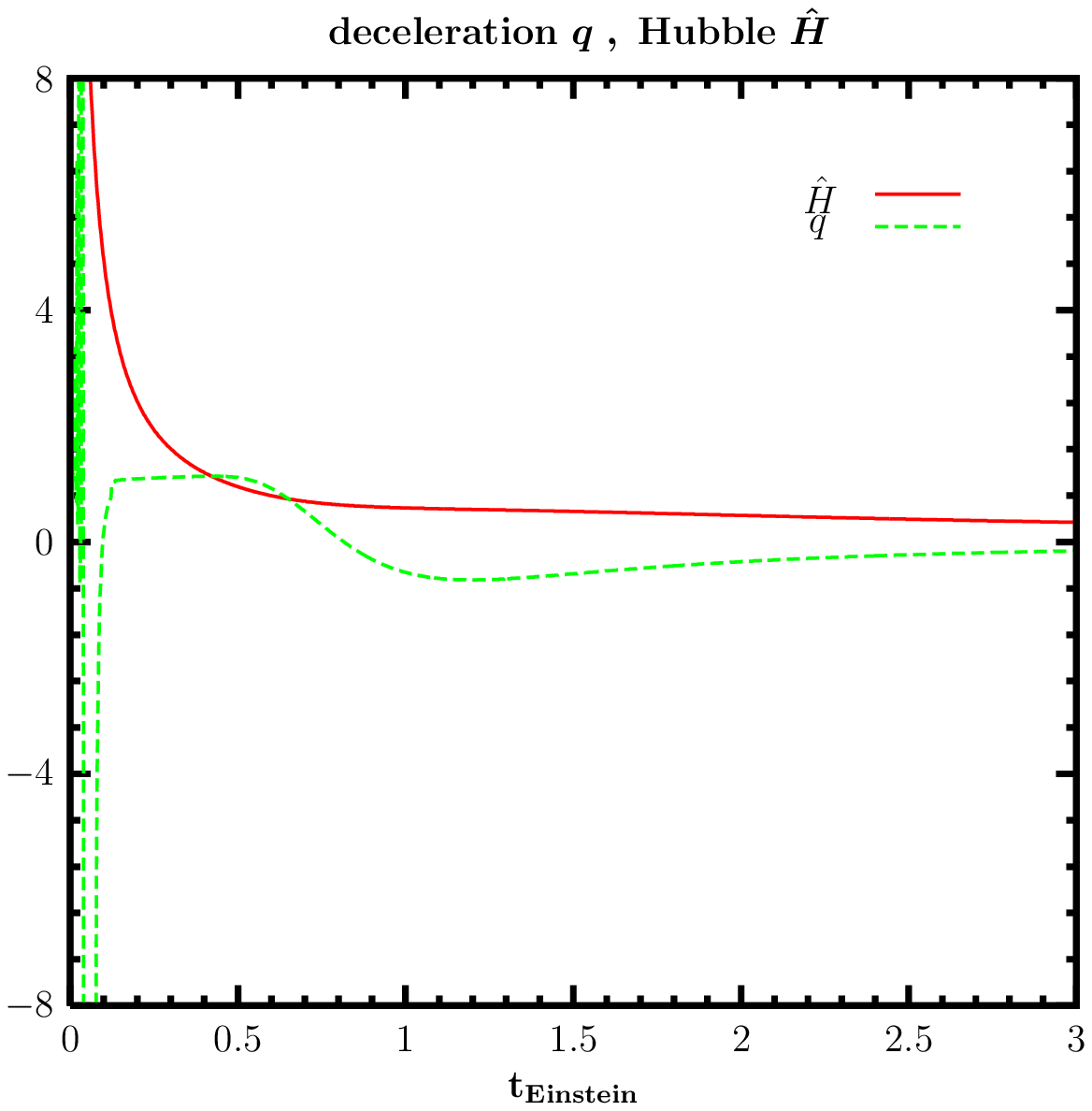}
\includegraphics[width=7cm]{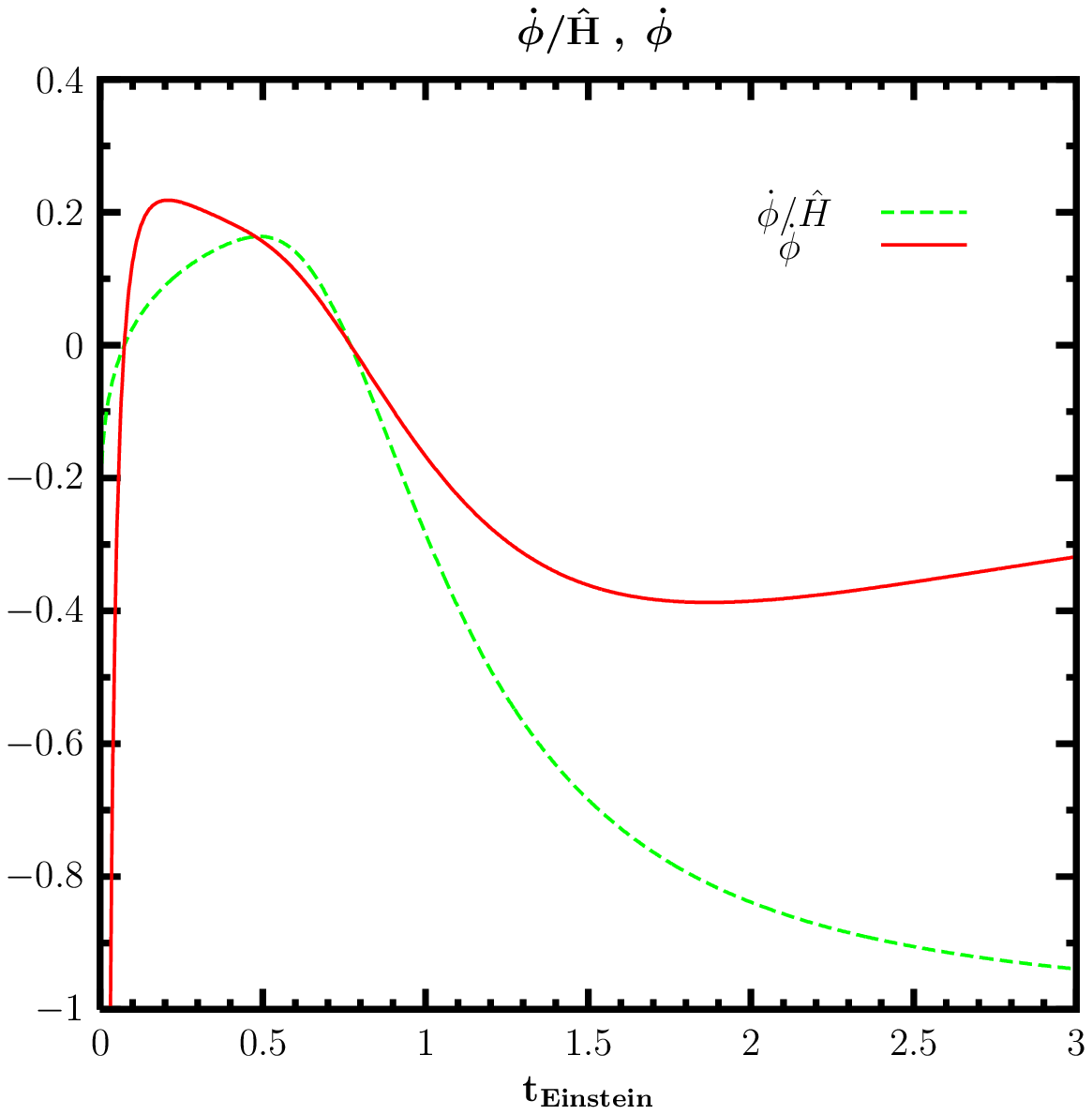}
\end{center}
\caption[]{ Left panel: The deceleration $q$ and the dimensionless
Hubble expansion rate $\hat{H}\equiv \frac{H}{\sqrt{3} H_0}$ as
functions of $t_{Einstein}$.  Right panel : The derivative of the
dilaton and its ratio to the dimensionless expansion rate.  }
\label{fig5}  
\end{figure}

With these inputs we can solve the
system of differential equations involved and follow the evolution of
all parameters of interest to past and future times. In figures
\ref{fig12} to \ref{fig5} we present sample outputs of our analysis for $\varrho_b=0.238,
\varrho_e=0.0$ and $w_e=0.5$. We have assumed that Dark Matter is contained 
in  $\varrho_b$ and the value $0.238$ for $\varrho_b$  corresponds to the observed 
central value 
$\Omega_{matter} h_0^2 = 0.127$, \cite{wmap}, with $h_0 =0.73$. The value of the exotic matter 
density today is assumed vanishing and  
the dilaton initial value is taken $\phi_0=0.0$ as well. 
In the left panel of \ref{fig12} we present the dilaton
$\phi$, the deficit $Q$ and the ratio $a/a_0$ of the cosmic scale
factor to its value today $a_0$ as functions of the Einstein time
$t_{Einstein}$. The present time is located where $a/a_0=1$ and in the
plot corresponds to $t_{today} \simeq 1.07$. The Bing Bang Singularity
(BBS) is located at the point $t_{Einstein}=0$. We have ignored at
this level inflation. If inflation dynamics is included the solutions 
are expected to change slightly when we approach the  
time Universe exited from the inflation period. 
Also the string dynamics plays a significant r\^ole near
BBS and on these grounds the solutions are expected to be modified
near the origin of time. In fact in such early regimes our 
$\sigma$-model and ${\cal O}(\alpha ')$ analysis breaks down.
Notice that the solutions obtained for these
inputs tend asymptotically in cosmic time~\cite{diamandis,emnw} 
to the conformal invariant solutions found 
in \cite{aben}. This is a rather general feature. 
In the specific case considered here, these asymptotic forms 
are obtained for times $\; t \geq 1.3$
that is greater than the present time $t_{today} \sim 1.07$.

In all cases, in the present era the
Universe has not reached, as yet, the asymptotic regime and the non-critical
terms play a significant r\^ole. On the right panel of 
figure \ref{fig12}
the values of $\Omega_i \equiv \rho_i/\rho_c$ for the various species, 
$i=b,r,e$ and $i=\phi$ for the dilaton,   
as functions of $t_{Einstein}$ are shown. For comparison, the contribution
of the non-critical terms to the Friedmann equation 
is also shown,  labelled by $\Omega_{noncr}$.  
Notice the behaviour of the exotic
matter density which is slightly negative for times
$t_{Einstein}>1.0\;$, ``dragged'' mainly by the action of the
non-critical terms.  

The bulk of the total energy today is carried by the dilaton whose 
energy is almost four times the energy carried out by the
dust and radiation together. This is more clearly seen on the left
panel of figure \ref{fig34} where the ratios of the dilaton, exotic
matter and non-critical densities to the dust and radiation energy
density are displayed. On the right panel of that figure the
quantities $\rho_{b} \;a^3$, for ''dust", $ \; \rho_r \;a^4$ and
$\rho_{e} \;a^2$ are shown. The dust matter density deviates
slightly from the law $\varrho_b = constant /a^3$ at certain epochs
due to its interaction with the dilaton. The exotic matter behaves
almost as $\varrho_e = constant /a^2$ asymptotically in agreement with
the asymptotic solutions considered in \cite{emnw}.

On the left panel of figure \ref{fig5} we present the
deceleration $q$ and the dimensionless Hubble expansion rate. The
deceleration today has been taken  $q_0=-0.61$ and stays negative
for future times. It is worth pointing out that the change from the
acceleration to the deceleration phase started at redshifts $z \approx
0.20$. This is a general characteristic  and not a specific feature of the 
particular sample outputs. For all cases the transition
to acceleration occurred at redshifts $z \sim {\cal{O}}(0.15-0.20)$. We shall 
return to this point later. 

On the right panel of figure \ref{fig5} 
we present the derivative of the dilaton value, which
governs the non-dissipative pressure term in the continuity equation
for matter and exotic matter, as well as its ratio to
$\hat{H}$. $\dot{\phi}$ behaves like $- 1/t_{Einstein}$ asymptotically tending to zero for very large 
times, not shown in the figure. Its ratio to
$\hat{H}$ tends to -1 for large times. Although we are not in the asymptotic regime for the
values shown in the figure this tendency is clearly seen. 
Recall the $\hat{H}$ is defined by 
$\hat{H} \equiv H / \sqrt{3} H_0$, that is its value today is $1/\sqrt{3}$.

The fact that this ratio is of
order of unity for certain epochs shows that it contributes on equal
footing with the expansion rate and cannot be neglected. In certain
epochs it acts constructively ( destructively ) with the expansion
rate depending whether it is negative ( positive ), tending to
decrease ( increase ) the energy density diluted by the
expansion. This is so because it appears with negative sign relative
to the Hubble rate term in the continuity equation (\ref{contin}).  This is
responsible for the behaviour of the dust density $\varrho_b$ for
large times, $t_{Einstein}> 1$, which decreases at a rate faster than
$a^{-3}$. Notice that for the exotic matter the effect of this term
may be reversed since the relative strength of $\dot{\phi}$ to
$\hat{H}$ is weighted by the factor $\frac{1-3 w_e}{1+w_e}$, as  be seen from 
(\ref{contin}), which
depending on the value of $w_e$ may be negative or positive.

\begin{figure}[H] 
\begin{center}
\includegraphics[width=7cm]{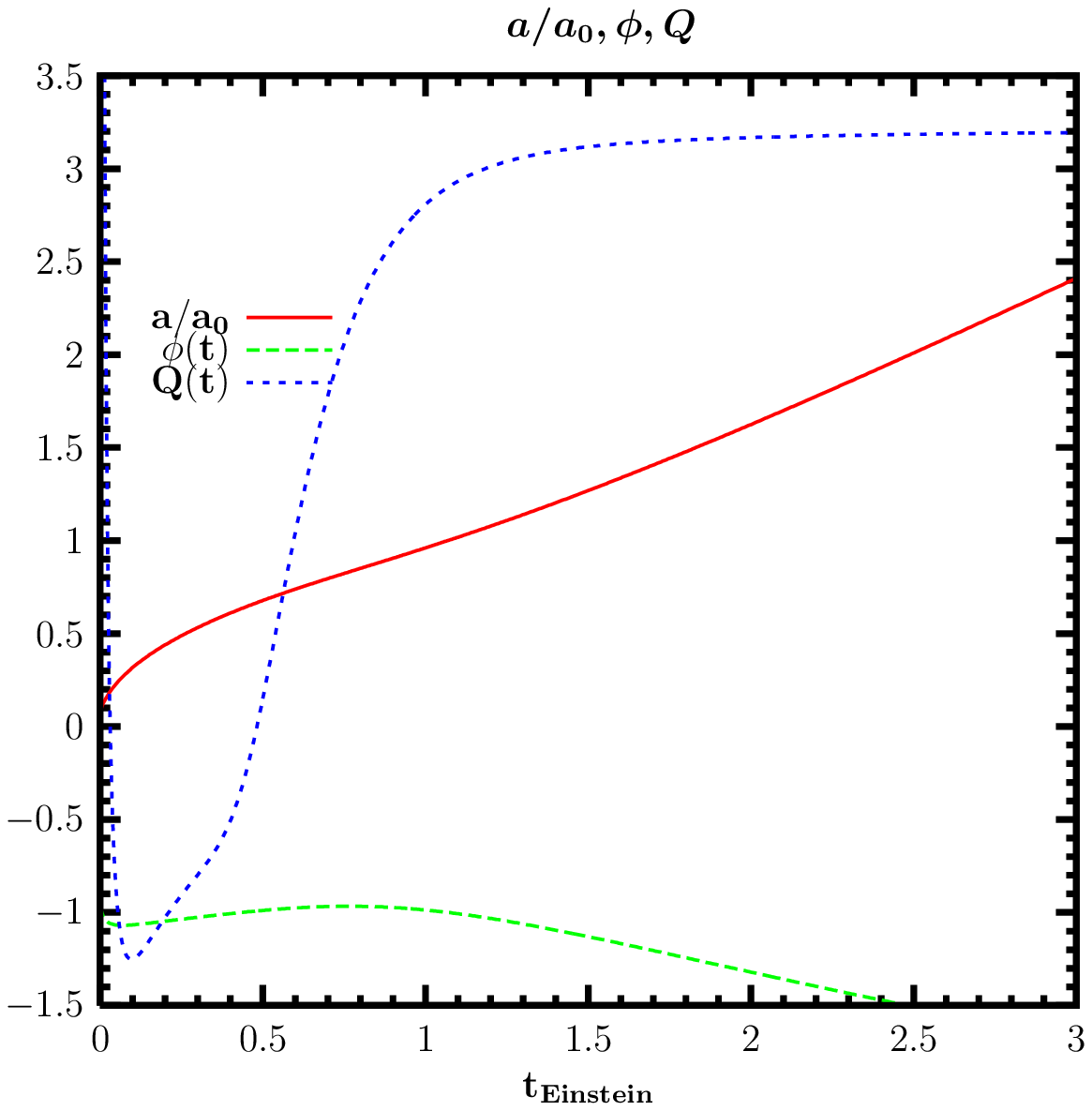}
\includegraphics[width=7cm]{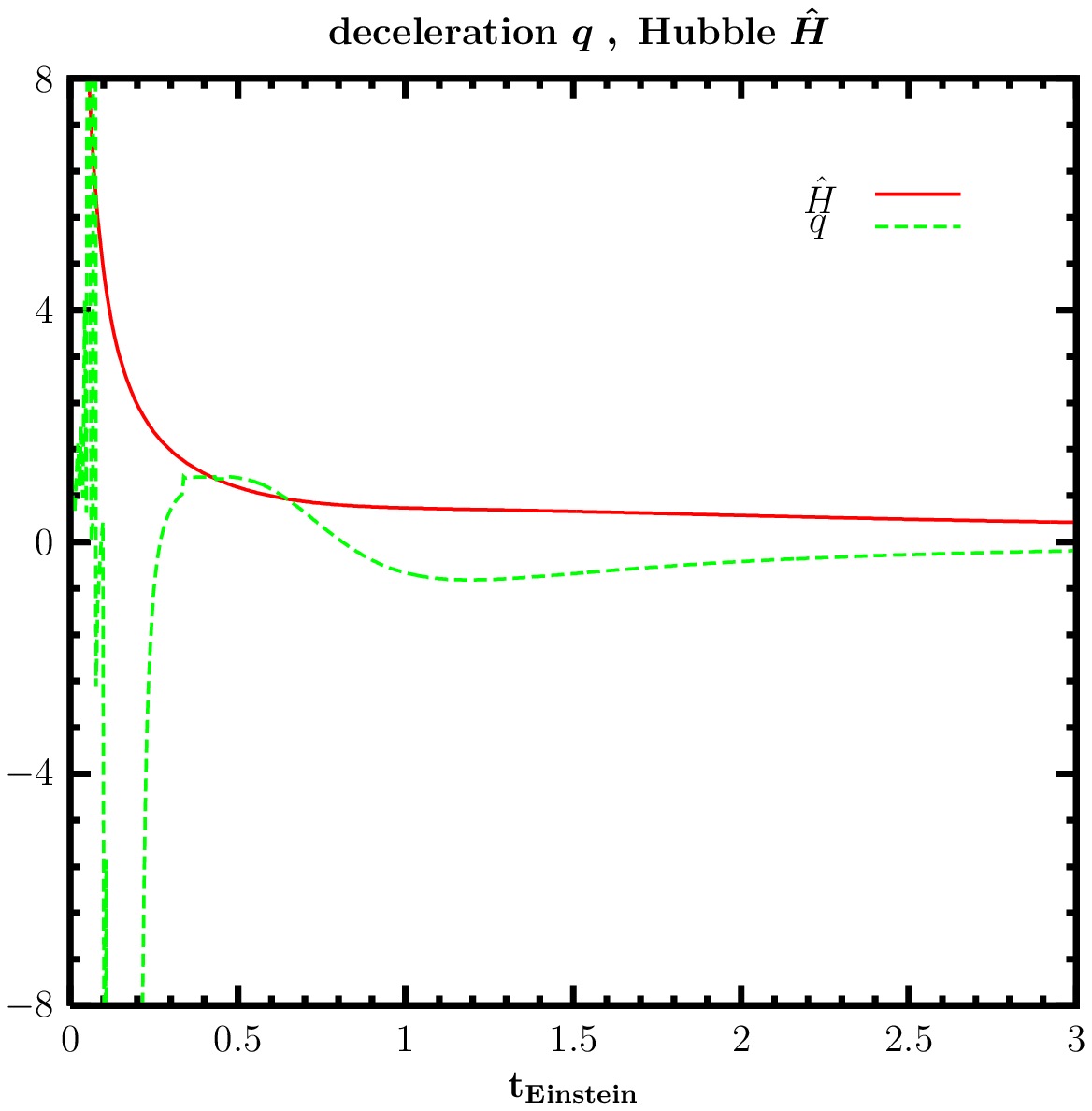}
\end{center}
\caption[]{ Left panel: The dilaton $\phi$, the (square root of the) 
central charge deficit
$Q$ and the ratio $a/a_0$ of the cosmic scale factor as functions of
the Einstein time $t_{Einstein}$. The inputs are as in figure
\ref{fig12} with only changing the dilaton to $\phi_0=-1.0$.  Right
panel: The deceleration and $\hat{H}$ for the same inputs.  }
\label{negfi}  
\end{figure}

For the same inputs and changing only the initial value of the dilaton
to $\phi_0=-1.0$ we display the dilaton, the cosmic scale factor and
the charge $Q$ in the left panel of figure \ref{negfi}. The entrance
into the asymptotic regime is delayed slightly due to the behaviour
of $Q$. For positive initial values for the dilaton field the
situation is altered. However for positive initial values the string
coupling becomes large and the perturbative solutions are not
valid. On the right panel of the same figure we display the
deceleration and the $\hat{H}$ for comparison with the case considered
previously. Notice that in this case too the deceleration behaves
almost in the same way.

Another important feature of the Q-cosmology is the relation of the 
present- and late -eras acceleration of the Universe to the 
string coupling $g_s=e^\phi$. In \cite{emnw}, 
based on the detailed, but matterless, Liouville model of \cite{diamandis},  
it has been argued that the 
acceleration of the Universe at late-eras, where matter
effects are suppressed, is proportioinal to the square of the 
string coupling
\begin{equation}
- q = g_s^2
\label{accelcoupl}
\end{equation} 
This seems to be a general feature of Liouville or Q-cosmology,
which asymptotically turn to the solutions of \cite{aben}.

\begin{figure}[H] 
\begin{center}
\includegraphics[width=7cm,height=7.16cm]{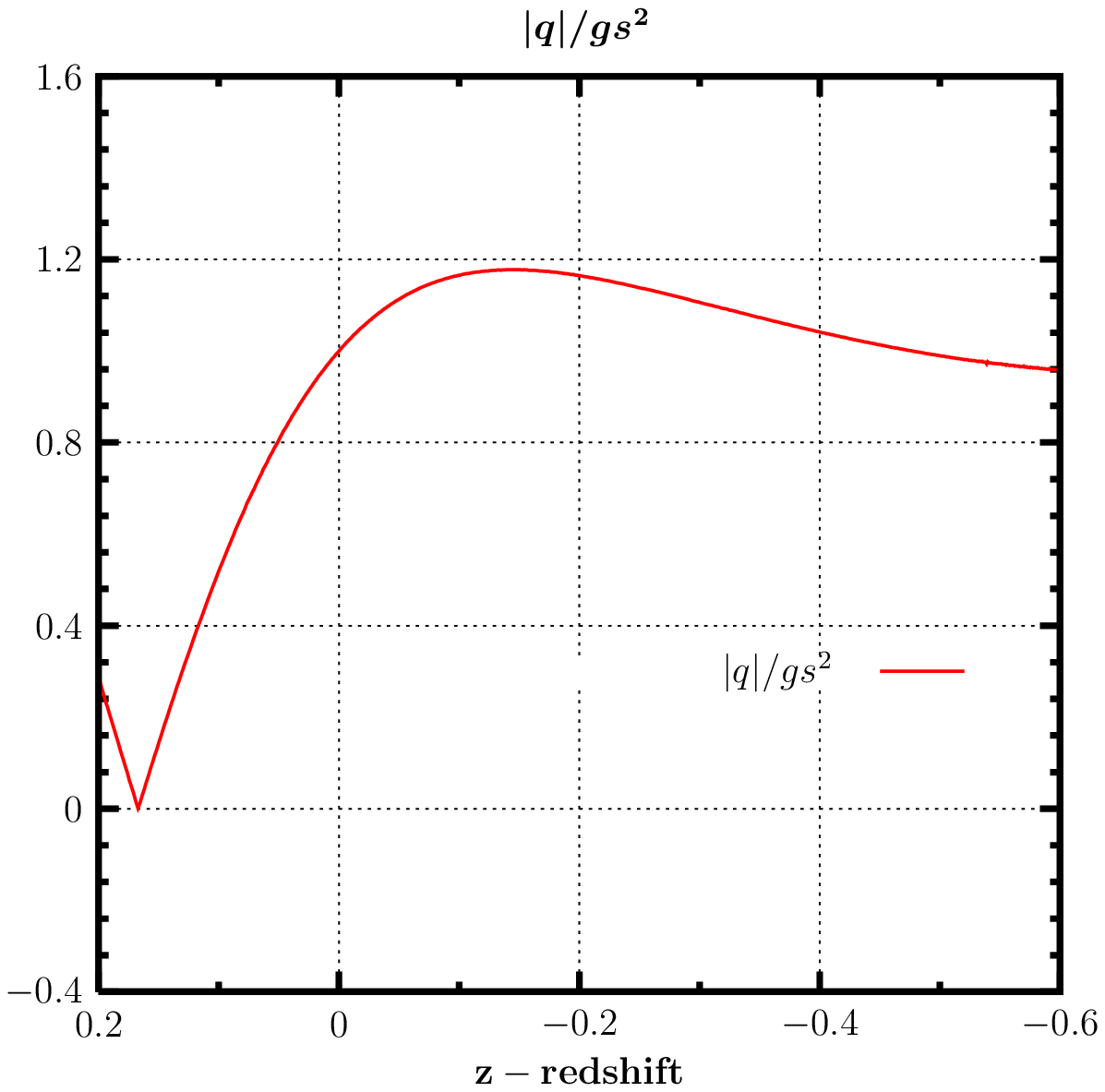}
\includegraphics[width=7cm]{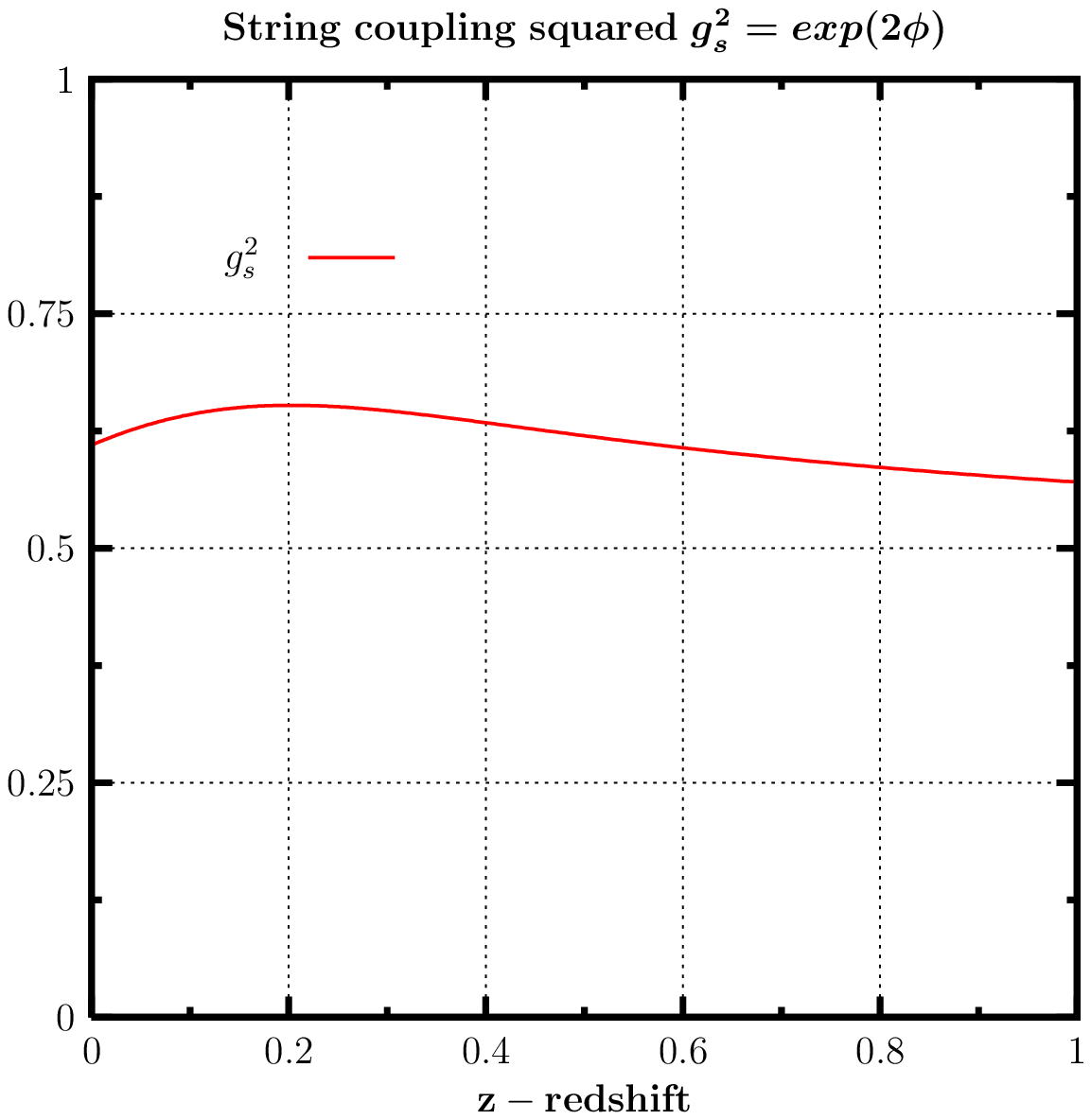}
\end{center}
\caption[]{ 
Left panel: 
The ratio $|q|/g_s^2$ as function of the redshift for $z$ ranging from $z=0.2$ 
to future values $z=-0.6$, for the inputs discussed in the main text. 
The rapid change near $z\approx 0.16$ signals the 
passage from deceleration to the acceleration period. 
Right panel: The values of the string coupling constant plotted 
versus redshift value in the range $z=0.0-1.0$. 
}
\label{qasz}  
\end{figure}

A detailed test of (\ref{accelcoupl}) is presented on the left panel of 
figure  \ref{qasz}. The input data are as in the previous figures with the 
exception of the dilaton whose value is taken $\approx 0.25$ so that 
$|q|/g_s^2$ is unity at present time. Notice that its value smoothly tends to 
unity for late eras too not deviating significantly from unity in intermediate 
epochs. The rapid change of this ratio, near redshift values $z \approx 0.16$, signals 
the change of the sign in $q$ from positive to negative values and hence the entrance 
to acceleration era. The string coupling constant 
is less than unity, as is shown in the right panel, which demonstrates that we are 
indeed within the perturbative regime of the string theory and our solutions are valid.

\begin{figure}[H] 
\begin{center}
\includegraphics[width=7cm,height=7.0cm]{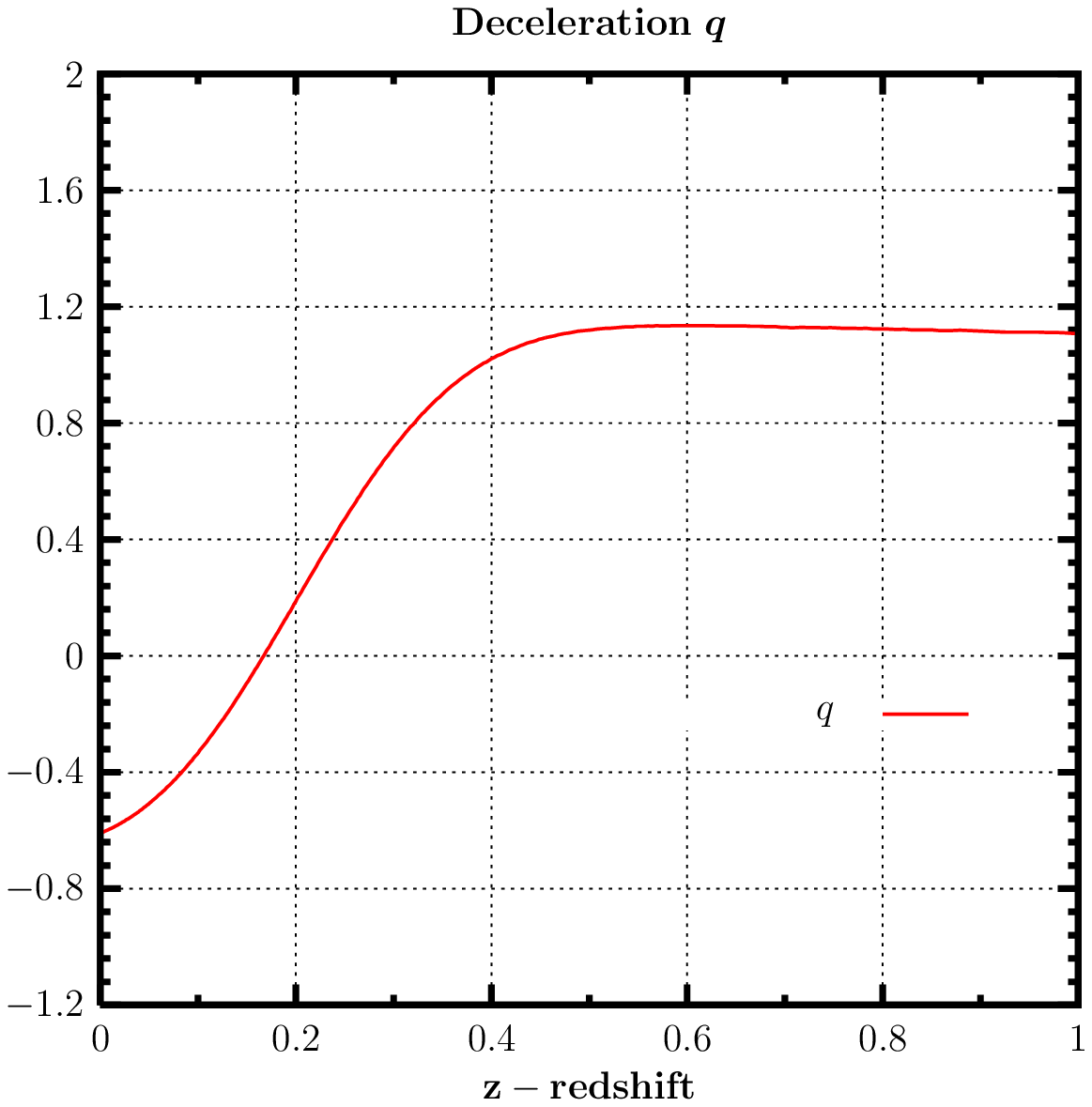}
\end{center}
\caption[]{ 
The deceleration as function of redshift values in the range $z=0.0-1.0$. 
The inputs are as in figure \ref{qasz}.
}
\label{decel}
\end{figure}

In figure \ref{decel} we display the deceleration as function of the redshift values in the 
range $0.0 - 1.0$ to cover the range of the high-z supernovae data which provides evidence 
on entrance to acceleration period at redshifts in the vicinity $z \approx 0.2$. 
The shape of the curve is strikingly similar to figure 7 of ref. \cite{Shapiro:2005nz}, in 
which a fit to SNeIa data is attempted using conventional cosmological models.  

With these remarks we conclude our analysis of this case study 
of Q-cosmology. The numerical solutions we have found are compatible 
in general terms with the current situation of an accelerating 
Universe, suggested by the astrophysical data. 
However, to complete the analysis it is necessary to 
perform detailed fits 
of these results to the actual astrophysical data, along the lines 
of \cite{mitsou}. 
From our analysis above it becomes clear that
the naive power-law behaviour 
in the redshift $(1 + z)$ (\ref{formulaforfit_txt}) 
of the various components appearing in the Hubble parameter $H(z)$,
which was used for late eras in the analysis of \cite{mitsou}
is not valid in the present model. 
More complex $z$-dependence appears to characterise $H(z)$ 
in our case. 
It remains to be seen 
whether the data favour this model, as compared with the 
other models in the literature. 

Moreover, it is also of outmost importance to study 
the effects of the off-shell and dilaton dissipative terms 
on the available parameter space of the 
relic densities in the context of supersymmetric
dark matter models in the $Q$-cosmology framework.
As discussed briefly in this work, it is expected that
the current constraints~\cite{nano} will change, rather significantly,
in view of the results of our analysis according to which in the 
present (and earlier eras) the effects of the off-shell Liouville terms
are important.  
This is left for future work \cite{reliclmn} .

%
%
%

\section*{Acknowledgements}

The work of G.A.D., B.C.G., A.B.L. and 
N.E.M.\ is partially supported by funds made available
by the European Social Fund (75\%) and National (Greek)
Resources (25\%) - ( EPEAEK~II ) PYTHAGORAS.
The work of D.V.N.\ is supported by D.O.E.\ grant
DE-FG03-95-ER-40917. N.E.M. wishes to thank the Department of Theoretical
Physics of the University of Valencia (Spain) for the
hospitality during the last stages of this work. 


\begin{thebibliography}{cc}

\bibitem{snIa} A.~G.~Riess \textit{et al.} [Supernova Search Team
Collaboration],
Astron.\ J.\ \textbf{116}, 1009 (1998) [arXiv:astro-ph/9805201];
B.~P.~Schmidt {\it et al.}  [Supernova Search Team Collaboration],
Astrophys.\ J.\  {\bf 507}, 46 (1998)
[arXiv:astro-ph/9805200];
S.~Perlmutter {\it et al.}  [Supernova Cosmology Project Collaboration],
Astrophys.\ J.\  {\bf 517}, 565 (1999)
[arXiv:astro-ph/9812133];
J.~P.~Blakeslee {\it et al.}  [Supernova Search Team Collaboration],
Astrophys.\ J.\  {\bf 589}, 693 (2003)
[arXiv:astro-ph/0302402];
A.~G.~Riess {\it et al.}  [Supernova Search Team Collaboration],
Astrophys.\ J.\  {\bf 560}, 49 (2001)
[arXiv:astro-ph/0104455];
  A.~G.~Riess {\it et al.}  [Supernova Search Team Collaboration],
Astrophys.\ J.\  {\bf 607} (2004) 665
[arXiv:astro-ph/0402512].

\bibitem{wmap}D.~N.~Spergel {\it et al.}  [WMAP Collaboration],
Astrophys.\ J.\ Suppl.\  {\bf 148}, 175 (2003)
[arXiv:astro-ph/0302209];
D.~N.~Spergel {\it et al.},
  arXiv:astro-ph/0603449.

\bibitem{Upadhye:2004hh}
  A.~Upadhye, M.~Ishak and P.~J.~Steinhardt,
  Phys.\ Rev.\ D {\bf 72} (2005) 063501
  [arXiv:astro-ph/0411803].

\bibitem{deceldata} A.~G.~Riess {\it et al.}  [Supernova Search Team Collaboration],
  Astrophys.\ J.\  {\bf 607}, 665 (2004)
  [arXiv:astro-ph/0402512];
N.~A.~Bahcall, J.~P.~Ostriker, S.~Perlmutter and P.~J.~Steinhardt,
  Science {\bf 284}, 1481 (1999)
  [arXiv:astro-ph/9906463].

\bibitem{aben}
I.~Antoniadis, C.~Bachas, J.~R.~Ellis and D.~V.~Nanopoulos,
  Phys.\ Lett.\ B {\bf 211}, 393 (1988);
Nucl.\ Phys.\ B {\bf 328}, 117 (1989);
Phys.\ Lett.\ B {\bf 257}, 278 (1991).

\bibitem{emnw} J.~R.~Ellis, N.~E.~Mavromatos and D.~V.~Nanopoulos,
Phys.\ Lett.\ B {\bf 619}, 17 (2005) [arXiv:hep-th/0412240]; 
J.~R.~Ellis, N.~E.~Mavromatos, D.~V.~Nanopoulos and
  M.~Westmuckett, 
Int.\ J.\ Mod.\ Phys.\ A {\bf 21}, 1379 (2006)
  [arXiv:gr-qc/0508105].  

\bibitem{polchinski} J.~Polchinski,
{\it String theory}, Vols. I \& II (Cambridge University Press, 1998).

\bibitem{diamandis} G.~A.~Diamandis, B.~C.~Georgalas, N.~E.~Mavromatos
and E.~Papantonopoulos,
Int.\ J.\ Mod.\ Phys.\ A {\bf 17}, 4567 (2002)
[arXiv:hep-th/0203241];
G.~A.~Diamandis, B.~C.~Georgalas, N.~E.~Mavromatos, E.~Papantonopoulos and I.~Pappa,
Int.\ J.\ Mod.\ Phys.\ A {\bf 17}, 2241 (2002)
[arXiv:hep-th/0107124].

\bibitem{ddk} F.~David,
Mod.\ Phys.\ Lett.\ A {\bf 3}, 1651 (1988);
J.~Distler and H.~Kawai,
Nucl.\ Phys.\ B {\bf 321}, 509 (1989);
J.~Distler, Z.~Hlousek and H.~Kawai,
Int.\ J.\ Mod.\ Phys.\ A {\bf 5}, 391 (1990);
see also: N.~E.~Mavromatos and J.~L.~Miramontes,
Mod.\ Phys.\ Lett.\ A {\bf 4}, 1847 (1989);
E.~D'Hoker and P.~S.~Kurzepa,
Mod.\ Phys.\ Lett.\ A {\bf 5}, 1411 (1990).



\bibitem{emn} J.~R.~Ellis, N.~E.~Mavromatos and D.~V.~Nanopoulos,
Phys.\ Lett.\ B {\bf 293}, 37 (1992)
[arXiv:hep-th/9207103];
  Mod.\ Phys.\ Lett.\ A {\bf 10}, 1685 (1995)
  [arXiv:hep-th/9503162].
{\it Invited review for the special Issue of J.\ Chaos Solitons Fractals},
Vol.\ 10, (eds. C. Castro amd M.S. El Naschie,
Elsevier Science, Pergamon 1999) 345
[arXiv:hep-th/9805120];

\bibitem{gsw} M.B. Green, J.H. Schwarz and E. Witten, {\it Superstring
Theory}, Vols. I \& II (Cambridge University Press, 1987).

\bibitem{gasperini} M.~Gasperini, F.~Piazza and G.~Veneziano,
  Phys.\ Rev.\ D {\bf 65}, 023508 (2002)
  [arXiv:gr-qc/0108016];
  T.~Damour and A.~M.~Polyakov,
  Nucl.\ Phys.\ B {\bf 423} (1994) 532
  [arXiv:hep-th/9401069].

\bibitem{mitsou} J.~R.~Ellis, N.~E.~Mavromatos, 
V.~A.~Mitsou and D.~V.~Nanopoulos,
  arXiv:astro-ph/0604272.

\bibitem{baryon} 
D.~J.~Eisenstein {\it et al.},
  Astrophys.\ J.\  {\bf 633} (2005) 560
  [arXiv:astro-ph/0501171];

S.~Cole {\it et al.}  [The 2dFGRS Collaboration],
  Mon.\ Not.\ Roy.\ Astron.\ Soc.\  {\bf 362}, 505 (2005)
  [arXiv:astro-ph/0501174];
for a critical comprehensive recent review see: E.~V.~Linder,
  arXiv:astro-ph/0507308 and references therein.

\bibitem{kaluza} M.~Minamitsuji, M.~Sasaki and D.~Langlois,
    Phys.\ Rev.\ D {\bf 71}, 084019 (2005)
    [arXiv:gr-qc/0501086].

\bibitem{curci} G.~Curci and G.~Paffuti,
    Nucl.\ Phys.\ B {\bf 286}, 399 (1987).

\bibitem{kolb} E.~W.~Kolb and M.~S.~Turner,
{\it The Early Universe},  
(Frontiers in physics, 69, Redwood City, USA: Addison-Wesley (1990)).  

\bibitem{nano}A.~B.~Lahanas, D.~V.~Nanopoulos and V.~C.~Spanos,
  Phys.\ Lett.\ B {\bf 518}, 94 (2001)
  [arXiv:hep-ph/0107151];
J.~R.~Ellis, K.~A.~Olive, Y.~Santoso and V.~C.~Spanos,
  Phys.\ Lett.\ B {\bf 565}, 176 (2003)
  [arXiv:hep-ph/0303043];
A.~B.~Lahanas and D.~V.~Nanopoulos,
  Phys.\ Lett.\ B {\bf 568}, 55 (2003)
  [arXiv:hep-ph/0303130];
for a comprehensive review see: A.~B.~Lahanas, N.~E.~Mavromatos and D.~V.~Nanopoulos,
  Int.\ J.\ Mod.\ Phys.\ D {\bf 12}, 1529 (2003)
  [arXiv:hep-ph/0308251] and references therein.

\bibitem{reliclmn} A.B.~Lahanas, N.E.~Mavromatos and D.V.~Nanopoulos
to appear.

\bibitem{Shapiro:2005nz}
  C.~Shapiro and M.~S.~Turner,
  arXiv:astro-ph/0512586.


\end{thebibliography}
\end{document}